\documentclass[letterpaper, inpress]{jds} 

\jdsmonth{July}                 
\jdsyear{2020}                  
\jdsvolume{xx}                  
\jdsissue{xx}                   
\jdsdoi{xx.xxxx/xxxxxxxxx}      
\jdsreceived{April, 2020}       
\jdsaccepted{May, 2020}         

\usepackage{amsfonts,amsmath,amssymb,amsthm}
\usepackage{booktabs}
\usepackage{graphicx}
\graphicspath{{./figures/}}
\usepackage{subfigure}
\usepackage{soul}
\usepackage{lipsum}
\newcommand{\mbold}[1]{\mbox{\boldmath ${#1}$}}
\usepackage{color}

\newcommand{\beginsupplement}{%
        \setcounter{table}{0}
        \renewcommand{\thetable}{S\arabic{table}}%
        \setcounter{figure}{0}
        \renewcommand{\thefigure}{S\arabic{figure}}%
     }

\usepackage{listings}
\usepackage{xcolor}
 \definecolor{codegreen}{rgb}{0,0.6,0}
\definecolor{codegray}{rgb}{0.5,0.5,0.5}
\definecolor{codepurple}{rgb}{0.58,0,0.82}
\definecolor{backcolour}{rgb}{0.95,0.95,0.92}
\lstdefinestyle{mystyle}{
    backgroundcolor=\color{backcolour},   
    commentstyle=\color{codegreen},
    keywordstyle=\color{magenta},
    numberstyle=\tiny\color{codegray},
    stringstyle=\color{codepurple},
    basicstyle=\ttfamily\footnotesize,
    breakatwhitespace=false,         
    breaklines=true,                 
    captionpos=b,                    
    keepspaces=true,                 
    numbers=left,                    
    numbersep=5pt,                  
    showspaces=false,                
    showstringspaces=false,
    showtabs=false,                  
    tabsize=2
}
\usepackage{booktabs}

\title[Bayesian Deep Learning Survival]{BDNNSurv: Bayesian Deep Neural Networks for Survival Analysis Using Pseudo Values}

\author[1]{Dai Feng\thanks{Corresponding author. Email: dai.feng@abbvie.com}}
\author[2]{Lili Zhao\footnote{Email: zhaolili@med.umich.edu}}
\affil[1]{Data and Statistical Sciences, AbbVie Inc., Illinois, USA}
\affil[2]{Department of Biostatistics, School of Public Health, University of Michigan, Ann Arbor, Michigan, USA}

\begin{document}

\maketitle

\begin{abstract}
There has been increasing interest in modeling survival data using deep learning methods in medical research. In this paper, we  proposed a Bayesian  hierarchical  deep  neural  networks  model  for modeling and prediction of survival data. Compared with previously studied methods, the new proposal can provide not only point estimate of survival probability but also quantification of the corresponding uncertainty, which can be of crucial importance in predictive  modeling and subsequent decision making. The favorable statistical properties of  point and uncertainty estimates were demonstrated by simulation studies and real data analysis. The Python code implementing the proposed approach was provided.\@
\end{abstract}


\section{Introduction}
\label{sec:intro}

Time-to-event (TTE) modeling is one of the most widely used
statistical analysis tools in health data science applications. TTE  analysis deals with outcomes that consist of the time to some event (death, occurrence of a disease, etc.). We refer this time as survival time hereafter. TTE  analysis is unique because the outcome of interest is not only whether an event has occurred or not (a binary outcome), but also when that event occurred (a continuous outcome). Furthermore, the outcome data maybe censored (i.e., incomplete observations), in particular right-censored; for example,  patients may be lost to follow-up without experiencing the event. Thus, traditional methods, such as logistic or linear regression, can not be used to  handle censored survival data.

Using neural networks to predict time-to-event has recently gained considerable attentions when there is a larger number of covariates \citep{Martinsson2016, Katzman:2018, Ching:2018, Zhu:2016, Gensheimer:2018, Fotso:2018, Lee:2018, Luck:2017, Giunchiglia:2018, zhao2019}. Neural networks is one of the most popular machine learning algorithms. It has been shown that neural networks outperform other algorithms in predictive modeling for certain tasks. The neural networks can outperform (1) widley used classical methods, e.g., 
Weibull regression \citep{Martinsson2016}, 
Cox proportional hazards model \citep{Katzman:2018, Gensheimer:2018, Fotso:2018, Lee:2018, Luck:2017, Giunchiglia:2018}, and (2) more recently developed statistical learning/machine learning methods, 
such as LASSO, ridge, minimax concave, and both $l_{1}$ and $l_2$ norm penalty \citep{Ching:2018, Zhu:2016, Giunchiglia:2018},
 random forest regression  \citep{Katzman:2018, Ching:2018, Zhu:2016, Lee:2018, Giunchiglia:2018}, 
 boosting based methods\citep{Ching:2018, Lee:2018}. 
 Moreover, the neural networks can handle not only high-dimensional but also unstructured data like medical images \citep{Zhu:2016}. In the TTE analysis, different types of neural networks, including deep neural networks (DNN), convolutional neural networks \textbf{\citep{Zhu:2016}} and recurrent neural networks \citep{Martinsson2016, Giunchiglia:2018} have been adopted.  
Besides different network architectures, various loss functions have been used based on the Weibull regression model \citep{Martinsson2016}, the Cox proportional hazards model \citep{Katzman:2018, Ching:2018, Zhu:2016} or  the discrete survival model \citep{Gensheimer:2018, Fotso:2018, Lee:2018, Luck:2017, Giunchiglia:2018, zhao2019}. 

The model built upon the Weibull regression assumes a parametric form of the distribution of survival time. The semi-parametric Cox proportional hazards regression instead studies effects of covariates on the hazards and relies on the proportional hazards (PH) assumption; that is,
the effect of a unit increase in a covariate is multiplicative with respect to the hazard rate. However, the quantity of direct interest sometimes is the survival probability rather than hazards. Furthermore, the PH assumption maybe questionable especially when the number of covariates is large, since every covariate needs to satisfy the PH assumption. Both parametric and semi-parametric models do not perform well when the  model is mis-specified.

 Recently, a flexible, simple deep neural network model was proposed in the survival setting \citep{zhao2019}. This approach consists of two steps. The first step is to compute the jackknife pseudo
survival probabilities in the discrete-time survival framework,
and substitute the survival times with censoring by these pseudo probabilities. By using the pseudo values, the analysis
for censored survival data is reduced to a regression problem
with a quantitative response variable, which greatly facilitates
the use of deep learning methods. In the second step, standard deep neural
networks can be directly applied, which avoids the difficulty
of designing a special cost function for the censored data. 

In this paper, we extended the work in \cite{zhao2019} to implement the two-step network model in a Bayesian framework. The framework enables quantification of the uncertainty of the predicted survival probability, which is an appealing feature in predictive modeling. The uncertainty is inherent in machine learning due to measurement error in data, randomness of estimated parameter values and model inadequacy. A key characteristic that distinguishes statistically rigorous approaches to prediction from other approaches is the ability to accurately quantify uncertainty of a prediction. The quantification of uncertainty can be harnessed to potential inspection of over-fitting, accurate comparison of treatment efficacy/effectiveness and identification of difficult diagnostic cases for further evaluation \citep{ghahramani2015,leibig2017, kendall2017, wager2014}. 

The remainder of this article is organized as follows. In
section \ref{sec:method}, we describe the details of our Bayesian deep neural networks model for survival analysis (BDNNSurv). In section \ref{sec:simul}, we present results from a simulation study to evaluate our proposed method. In section \ref{sec:real}, we demonstrate the application of our method to a prospective multicenter cohort study for cardiovascular disease. We conclude this
paper with discussions in section \ref{sec:final}.



\section{Methodology}
\label{sec:method}
Following our previous work in \cite{zhao2019},  we substituted the survival times with pseudo survival probabilities and then implemented the standard deep neural networks model for a regression problem with a quantitative outcome. In this section, we first briefly introduce the pseudo values calculation and then describe the BDNNSurv approach.

Let $T^*_i$ be the survival time for subject $i$, $C_i$ be the censoring time,
$T_i=\min(T^*_i,C_i)$ be the observed survival time, and $\delta_i=I(T^*_i \leq C_i)$ be the censoring
indicator. We assumed that survival times and censoring times are independent. Let $\mathbf{X_i}=(x_{i1},\cdots, x_{id})$ denote the $d$-dimensional covariates.  The pseudo-observations approach \citep{Andersen:2003,Klein:2005,Klein:2007,Andersen:2010,Tayob:2017,Xiang:2012,ZhaoMurray:2020} provides an efficient and straightforward way to study the association between the covariates and survival outcome in the presence of censoring.

Note that due to censoring, the survival probability at a specific time $t$ cannot be directly modeled as a binary outcome $I(T_i> t )$ on $\mathbf{X}_i$ ($i=1,\cdots,n$) using a generalized
linear model (GLM) with a \textit{logit} link, In the presence of
censoring, $I(T_i> t)$ is not observed for all subjects. In this case the Kaplan-Meier (KM)
estimator can be used to estimate the survival probability at any given time point. 
Based on the \textit{jackknife}
idea, a pseudo survival probability is computed for each (censored and uncensored) subject. For the
$i^{th}$ subject, the pseudo survival probability is computed by

\begin{align}
\widehat{S}_i(t) &= n \widehat{S}(t) - (n-1) \widehat{S}^{-i}(t),
\label{eqn:pseudo}
\end{align}
\noindent where  $\widehat{S}(t)$ is the KM estimator of $S(t)$ using all $n$ subjects and
$\widehat{S}^{-i}(t)$ is the KM estimate using sample size of $n-1$ by eliminating the $i^{th}$
subject. Then  $\widehat{S}_i(t)$ ($i=1,\cdots,n$) are used as a continuous response variable in the
standard regression analysis, regardless of censoring or not.

\begin{figure}[!h]
\vspace{-5cm}
\hspace{-2cm}
 \includegraphics[scale=0.8]{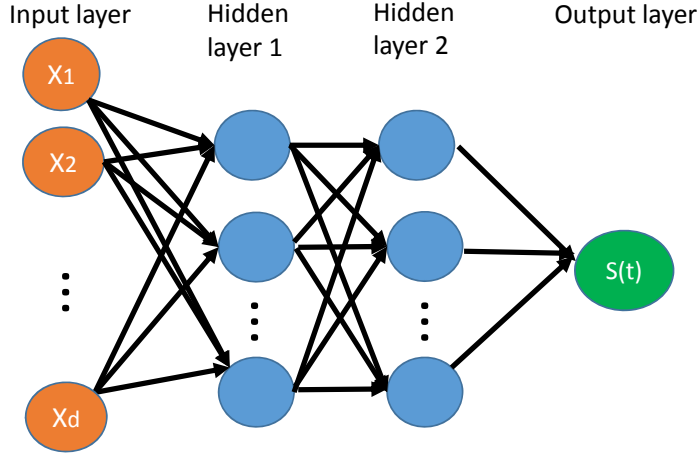}
 \vspace{-5cm}
\caption{A neural network with two fully connected hidden layers}
\label{fig:network}
\end{figure}

We applied the deep neural network to model the pseudo survival probability. Let $l$ be a layer of a deep neural network, $l=1, \ldots, L$; $d$ be the dimension of input layer; $N_l$ be the number of neurons in  layer $l$; $g_l: \mathbb{R}\rightarrow \mathbb{R}$ be an activation function; $A_l: \mathbb{R}^{N_{l-1}}\rightarrow \mathbb{R}^{N_l}$ be an affine linear map defined by a matrix $W_l \in \mathbb{R}^{N_{l-1}\times N_l}$ and an affine part $b_l \in \mathbb{R}^{N_{l}}$ via $A_l(\mathbf{X^*})=W_l\mathbf{X^*} + b_l$. Then $f: \mathbb{R}^{d}\rightarrow \mathbb{R}^{N_L}$  given by
\[f(\mathbf{X})=g_LA_L(g_{L-1}A_{L-1}(\ldots g_1(A_1(\mathbf{X}))), \quad \mathbf{X} \in \mathbb{R}^{d}\]
is a deep neural network (DNN). The input layer is layer 0 with $N_0=d,$ and the output layer has one dimension with $N_L=1$. The neural network is a function that maps input covariates $\mathbf{X}$ to a desired target value $f(\mathbf{X})$, the corresponding pseudo survival probability. When $l=1, f(\mathbf{X})=g_1A_1(\mathbf{X})$. Figure \ref{fig:network} shows an architecture of a neural network with two fully connected hidden layers.

To build a DNN, the architecture, including $L, N_l, g_l, l=1, \ldots, L$, are pre-determined by using, for example cross-validation. The parameters of affine transformation $W_l, b_l$ are traditonally obtained by minimizing the loss function or the objective function. For example, the loss function in \cite{zhao2019} for the numeric response of pseudo probabilities is the mean squared error. 
The optimization(minimization) is done using gradient descent. A fast algorithm for computing gradients is backpropagation \citep{nielsen2015}.

\begin{figure}[h]
\hspace{-3cm}
 \includegraphics[scale=0.8]{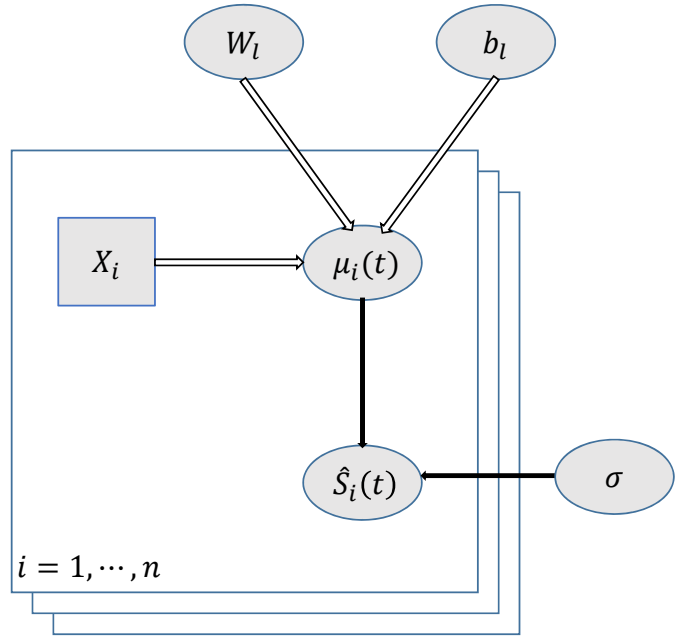}
\vspace{-5cm}
\caption{A directed graphical model representation of the BDNNSurv.}
\label{fig:dag}
\end{figure}

In the Bayesian hierarchical DNN model, the survival probability at each time $t$, $\widehat{S}_i(t)$, follows a normal distribution:
\begin{equation}
\label{eqn:bayes}
\widehat{S}_{i}(t) \sim N(\mbold{\mu_i(t)}, \sigma^2).
\end{equation}
The mean of the normal distribution, $\mbold{\mu_i(t)}$, is the transformation of covariates $\mathbf{X_i}$ to a desired target value $f_t(\mathbf{X_i})$, and the variance of the normal distribution is $\sigma^2$. For prior distributions of parameters of affine transformation $W_l, b_l$ in $f(\mathbf{X})$, we assumed:
the components of $W_l$, $W^c_l$, follow a normal distribution 
\begin{equation}
\label{eqn:priorW}
W^c_l \sim N(\mu_{W_l}, \sigma_{W_l}^2),\end{equation} 
and the components of $b_l$, $b_l^c$, follow a normal distribution 
\begin{equation} 
\label{eqn:priorb}
b_l^c \sim N(\mu_{b_l}, \sigma_{b_l}^2).
\end{equation}
Furthermore, we assumed $\sigma$ follows a uniform distribution $U(0, \sigma_{\textrm{max}})$. A directed acyclic graph (DAG) representation of the BDNNSurv model was shown in Figure \ref{fig:dag}.

The setup of the BDNNSurv, including the likelihood function in equation (\ref{eqn:bayes}) and priors in equations (\ref{eqn:priorW}) and (\ref{eqn:priorb}), is standard in  Bayesian hierarchical models. However, the automated Markov chain Monte Carlo (MCMC) approach can fail due to the lack of conjugate priors and high-dimensional parameter space in DNN.

A faster alternative to the MCMC is to use the variational inference (VI) \citep{wainwright2008}, which has been successfully applied in many large-scale problems \citep{hoffman2013, blei2017}. The VI turns the problem of computing a posterior distribution into an optimization problem. The traditional  VI algorithm requires tedious model-specific derivations and implementation, which hindered its wide usage. Recently an automate process of deriving scalable VI algorithms, the so-called automatic
differentiation variational inference (ADVI), has been developed in probabilistic programming \citep{kucukelbir2017}. The ADVI automatically develops an algorithm given a model, enabling an efficiently computation of Monte Carlo approximations. For more details on the ADVI, see \cite{kucukelbir2017} and references therein.

To implement the BDNNSurv, we adopted the same architecture, the good performance of which was demonstrated in  \cite{zhao2019}. There are three layers in the DNN with $N_1=8, N_2=4$, and $N_3=1$. The activation functions $g_1$ and $g_2$ are hyperbolic tangent functions and $g_3$ is sigmoid function. For the prior distributions of $W^c_l$ and $b^c_l$, we let  $\mu_{W_l}=\mu_{b_l}=0$ and $\sigma_{W_l}^2=\sigma_{b_l}^2=1$ for every layer $l=1,2,3$ and every component of $W$ and $b$. In addition, we assign an uniform prior $U (0, 0.2)$ as the prior distribution of standard deviation $\sigma$ of  $\widehat{S}_i(t)$.  Based on the posterior distributions of $W_l$ and $b_l$, we obtain  the posterior distributions of pseudo survival probabilities.

To adopt the ADVI, we used PyMC3 \citep{salvatier2016}, an open source probabilistic programming framework written in Python that builds on Theano \citep{theano2016} with ADVI implementation. The initial values of $W^c_l$ and $b^c_l$ were obtained from running the DNN model proposed in \cite{zhao2019}. The Python code implementing the proposed approach was included in the Supplementary Material.

KM estimates used in creating the pseudo values are subject to covariate-dependent censoring bias.
In this case, we propose to use the inverse of probability of censoring weighted (IPCW) estimator
for the survival function, denoted by $\hat{S}^W(t)$, which has been successful at reducing the
bias (\cite{Binder:2014,Xiang:2012,zhao2019, zhao2020}). We replace $\hat{S}(t)$ by $\hat{S}^W(t)$ in (\ref{eqn:pseudo})
to compute the IPCW pseudo conditional survival probabilities by

\begin{equation}
\hat{S}^W_{ij}(t_{j+1}|R_{j}) = R_{j} \hat{S}^W(t_{j+1}|R_{j}) - (R_{j}-1) \hat{S}^{{W}^{-i}}(t_{j+1}|R_{j}),
\label{pseudosw}
\end{equation}
where $\hat{S}^W(t) = \exp \{-\hat{\Lambda}^W(t)\},$ and $\hat{\Lambda}^W(t)$ is the IPCW
Nelson-Aalen estimator for the the cumulative hazard function and is estimated by
\begin{equation}
\hat{\Lambda}^W(t)=\sum^n_{i=1} \int^t_0 \frac{d N_i(u)  \hat{W_i}(u)}{\sum^n_{j=1} Y_j(u)\hat{W_j}(u)},
\label{eqn:NA}
\end{equation}
where $N_i(u)=I(T_i \leq u, \delta_i=1)$ is the observable counting process for subject
$i,$ $Y_j (u) = 1(T_j \geq u)$ is the at risk process for subject $j$, and $\hat{W_i}(u)$ is the
inverse of probability of censoring for subject $i$ at time $u$. By assigning different weights for
subjects based on their covariate values, $\hat{S}^W(t)$ is approximately unbiased if the censoring
distribution is correctly specified \citep{Binder:2014}.

To accurately model the censoring times as a function of covariates, we followed the method in \cite{zhao2020}. The random survival forest (RSF) proposed in \cite{ishwaran2008} was first applied to the full data to estimate the probability of censoring for each subject at each observed event time to compute $\hat{W_i}(u)$ in (\ref{eqn:NA}), and then these estimates were re-used in computing pseudo observations, which is the
same as the idea of re-using censoring regression coefficients from the
full data analysis in the leave-one-out estimator in \cite{Binder:2014}. We denote this method as BDNNSurv-IPCW hereafter.

\section{Simulation}
\label{sec:simul}

We conducted a simulation study to evaluate the performance of the newly proposed BDNNSurv approach. Besides the BDNNSurv, we adopted another approach based on the Bayesian Additive Regression Trees (BART) method in \cite{chipman2010} and compared the results from two approaches. BART is a flexible Bayesian non-linear regression approach and has been demonstrated to be competitive with widely used machine learning models. BART is formulated in a Bayesian hierarchical modeling framework, and as such provides both predictions and prediction interval estimates for a variety of quantities of interest. The original BART was extended to handle survival data using discrete-time survival analysis \citep{sparapani2016}.
Given the $k$ distinct event and censoring times: $0 < t_{(1)} < \ldots < t_{(k)} < \infty$, where $t_{(j)}$ is the $j$th order statistic among distinct observation times, the binary outcome of a subject $i$ with/without an event/censoring until time $t_{j}$ is modeled using package \emph{BART} \citep{rbart} in R. We used the default setup in the package except for the thinning parameter and coarsen times. We used thinning by a factor of 1 instead of 10 due to low computational speed of the BART as detailed in Section \ref{sec:final}. To overcome memory overflow, we used coarsen times per the quantiles $1/200, 2/200, \ldots, 1$.

In addition to BART, we compared the BDNNSurv with another Bayesian non-parametric, non-proportional hazards survival model proposed in \cite{de2009}, where 
an unconstrained model for survival regression was constructed based on the covariate-dependent Dirichlet process (DDP). The dependence across random distributions was modeled in an ANOVA-type fashion.  
We implemented the ANOVA DDP approach using package \emph{spBayesSurv} \citep{rddp} in R with the same parameters as specified in the example of the package.

Survival times were simulated from a Cox proportional hazards model with baseline hazard from log-logistic function using the inverse  cumulative distribution function method \citep{bender2005}. First, a variable $u$ was generated from a uniform(0, 1) distribution. Second, ten continuous covariates: $x_1, x_2, \ldots, x_{10}$, were independently generated from standard normal distributions; one binary covariate $z_1$ was generated from a Bernoulli distribution with probability taking 1 equal to 0.5. Let $X=(x_1, x_2, x_3)$. Finally, the survival time was equal to:
\begin{equation}
b\left\{\exp\left[-\log{(u)}\exp\left(-\beta f(X)\right)\right]-1\right\}^{(1/a)},
\label{eqn:sim}
\end{equation}
where \begin{equation}f(X)=\exp{\left(X^T {\bf V} X\right)},\label{eqn:fx}\end{equation}
\[{\bf V}=0.05\times\left(\begin{matrix} 
1 & \rho & \rho^2 \\
\rho & 1 & \rho\\
\rho^2 & \rho & 1
\end{matrix}\right),\]
$a=1.1, b=0.8$, $\beta=0.1$, and \textbf{$\rho=0.5$}. To simulate survival times, we introduced a non-linear function with interaction into the hazard function through $f(X)$. Furthermore, we simulated the covariates $x_4, x_5, \ldots, x_{10}$ and $z_1$ as noise variables. The noise variables were not used to generate survival times. However, they were included in the model.

In addition to survival times, the censoring times were independently generated from an exponential distribution. A rate parameter was chosen to obtain a censoring rate of about 20\%. Survival times greater than 15 were censored at 15. 

We generated one hundred data sets with each having 5000 subjects. Of the 5000 subjects, 75\% were randomly selected as training data and the remaining 25\% as test data. We estimated parameters using the training data and then predicted the survival probabilities along with corresponding 95\% credible intervals on the test data from the $10^{th}$ to $90^{th}$ quantile of the survival function, with an increment of 10\%. Note that the credible interval was used to quantify the uncertainty of survival probability. There is 95\% probability that a given credible interval contains the true value of corresponding survival probability.

To demonstrate the performance of BDNNSurv, we  investigated the cases with different sample sizes, censoring rates, numbers of predictor variables, and magnitude of parameter in matrix ${\bf V}$ of equation (\ref{eqn:fx}). Furthermore, we simulated data from a non-proportional hazards (non-PH) model. Finally, we considered the case of covariate-dependent censoring. See Table \ref{tab:sim} for different simulation setups.

To simulate data from a non-PH model, we let $a = a+1\times z_2$, where $z_2$ follows a Bernoulli distribution with the probability parameter being equal to 0.5. In the covariate-dependent censoring case, the censoring distribution was the same as the survival time distribution, except that we adjusted parameter $b$ to achieve 40\% censoring rate.

\begin{table}[h!]
\begin{center}
\caption{Different Simulation Setups}
\label{tab:sim}
\begin{tabular}{ c c c c c c c}\toprule
Case & Model & $N$ & Number of Covariates \textbf{X} & $\rho$ in $V$ & Censoring Rate \\\midrule

1 &	PH&  5000&		10&	0.5 & 20\%\\
2&	PH&	5000&		10&	0.5 & 40\%\\
3&	PH&	2000&		10&	0.5 & 20\%\\
4&	PH&	5000&		10&	0.95 & 20\%\\
5&	PH&	5000&		50&	0.5 & 20\%\\
6&	Non-PH&	5000&	10&	0.5 & 20\%\\
7&	PH IPCW&	5000&	10&	0.5 & 40\%\\\bottomrule
\end{tabular}
\end{center}
\end{table}

To compare different methods, we evaluated the statistical properties of both point and corresponding uncertainty estimates. For point estimates, we compared the bias and square root of mean squared error (MSE) of the estimated survival probabilities at different time points. The best model should have the smallest bias and MSE.  The coverage probabilities were calculated for 95\% credible intervals. The best model should have the coverage cloest to the nominal value 95\%. We compared our BDNNSurv to BART and ANOVA DDP. We also compared with DNNSurv \citep{zhao2019}, regarding the bias and squared MSE (DNNSurv does not provide estimate of uncertainty). In case 7, we also compared with the BDNNSurv-IPCW.

The simulation results of case 1 and 5-7 are shown in Figure \ref{fig:sim} and results of cases 2-4 are provided in Supplementary Material. For the point estimates, the BDNNSurv and DNNSurv had very similar results. In general, the BDNNSurv outperformed BART and ANOVA DDP, as evidenced by smaller bias, comparable MSE, and coverage probability closer to the nominal value. BART seems to perform the worst in most scenarios. ANOVA DDP didn't perform well when the number of noise variables is large (case 5), and when the covariate-dependent censoring is present (case 7). As expected, the BDNNSurv-IPCW outperformed the other methods, when the covariate-dependent censoring is present, because other methods can only handle independent censoring.

\begin{figure}
\centering
  \includegraphics[scale=0.22]{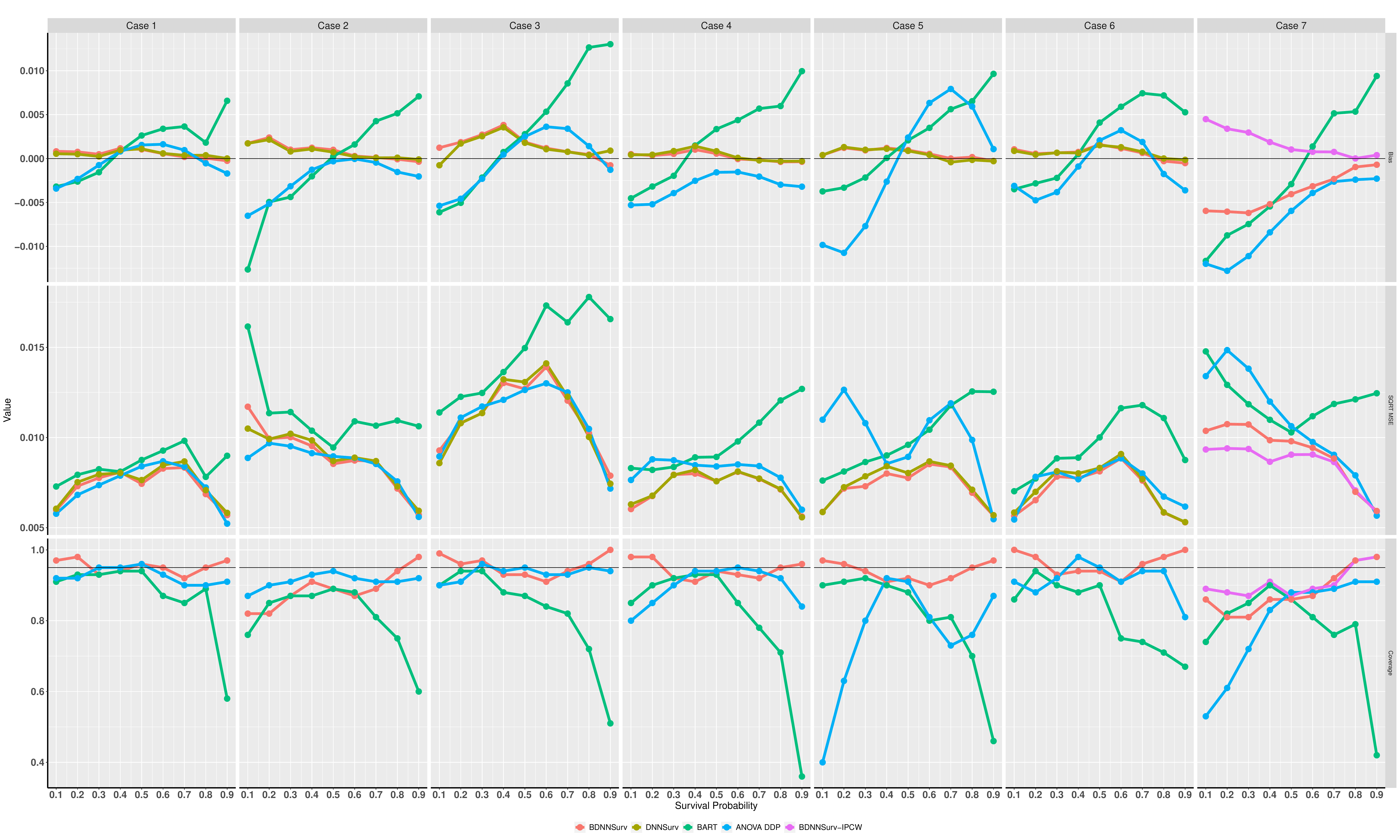}  
 \caption{The comparison of bias, $\sqrt{\textrm{MSE}}$, and coverage among results from BDNNSurv, BART, and ANOVA DDP using simulated data.}
\label{fig:sim}
\end{figure}

\section{Real Data}
\label{sec:real}

We applied the BDNNSurv approach to the  Cardiovascular Health Study (CHS) data. CHS is a prospective, multicenter cohort study for coronary heart disease and stroke. The study is sponsored by the National Heart Lung and Blood Institute of the National Institutes of Health.

The study was initiated in 1987 to determine the risk factors for
development and progression of cardiovascular disease (CVD) in older adults, with an emphasis on subclinical measures. The event of interest was time to CVD. The study has collected a large number of variables at baseline, including demographics (e.g., age, gender and race), family history of CVD, lab results, and medication information, with the goal to identity important risk factors for the CVD event. Detailed description of the study can be found in \cite{DrPhilos1993}. 

We selected 29 predictor variables after excluding the ones with more than 20\%  missing values. Furthermore, after excluding subjects with missing data in any of the selected predictor variables, we chose $5,380$ subjects, 65.2\% of which had CVD during the study period. We randomly selected 75\% subjects as training data and the other 25\% as test data. 

\begin{figure}
\centering
\begin{subfigure}
  \centering
  \includegraphics[scale=0.46]{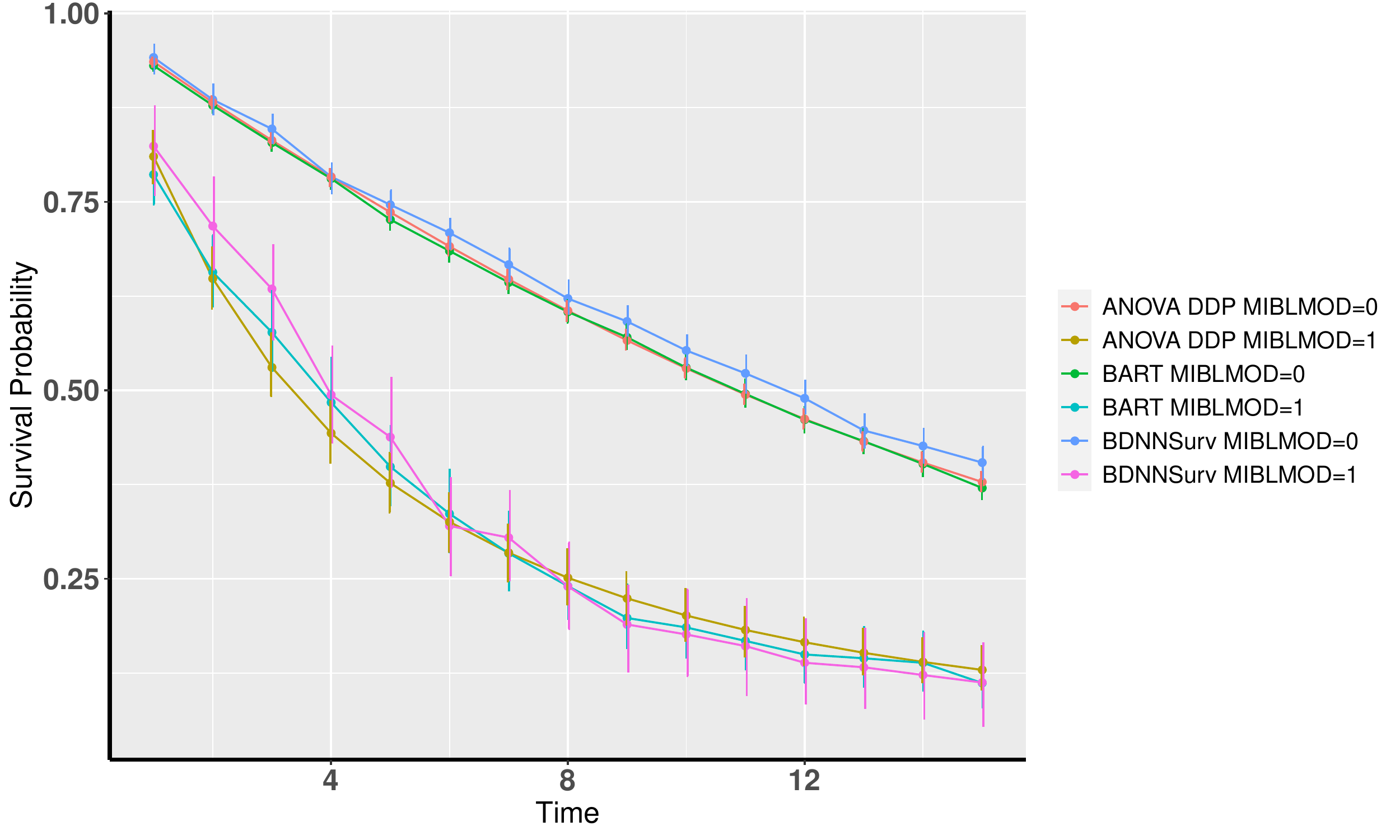}  
  \label{fig:sub-first}
\end{subfigure}
\begin{subfigure}
  \centering
  \includegraphics[scale=0.46]{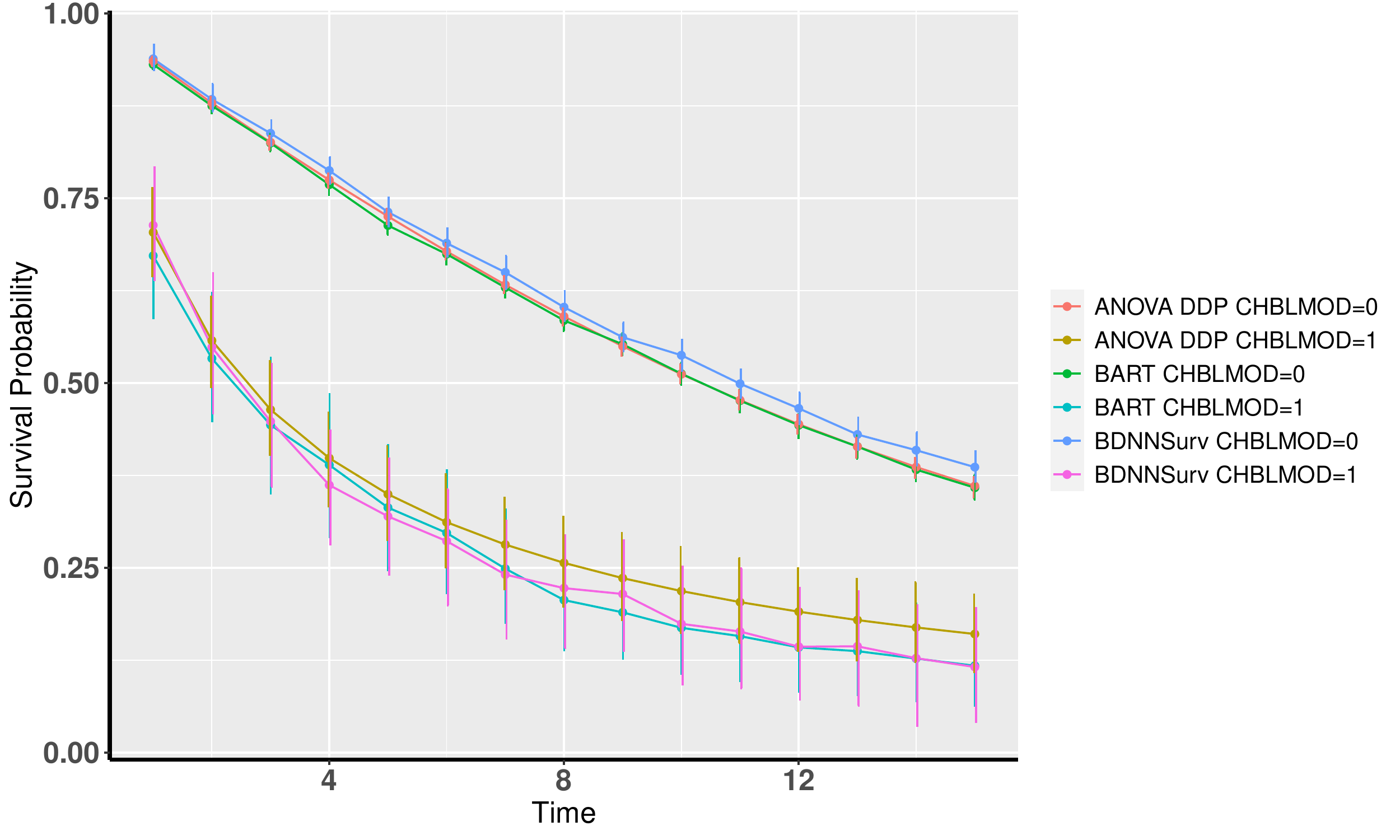}  
  \label{fig:sub-first}
\end{subfigure}
\begin{subfigure}
  \centering
  \includegraphics[scale=0.46]{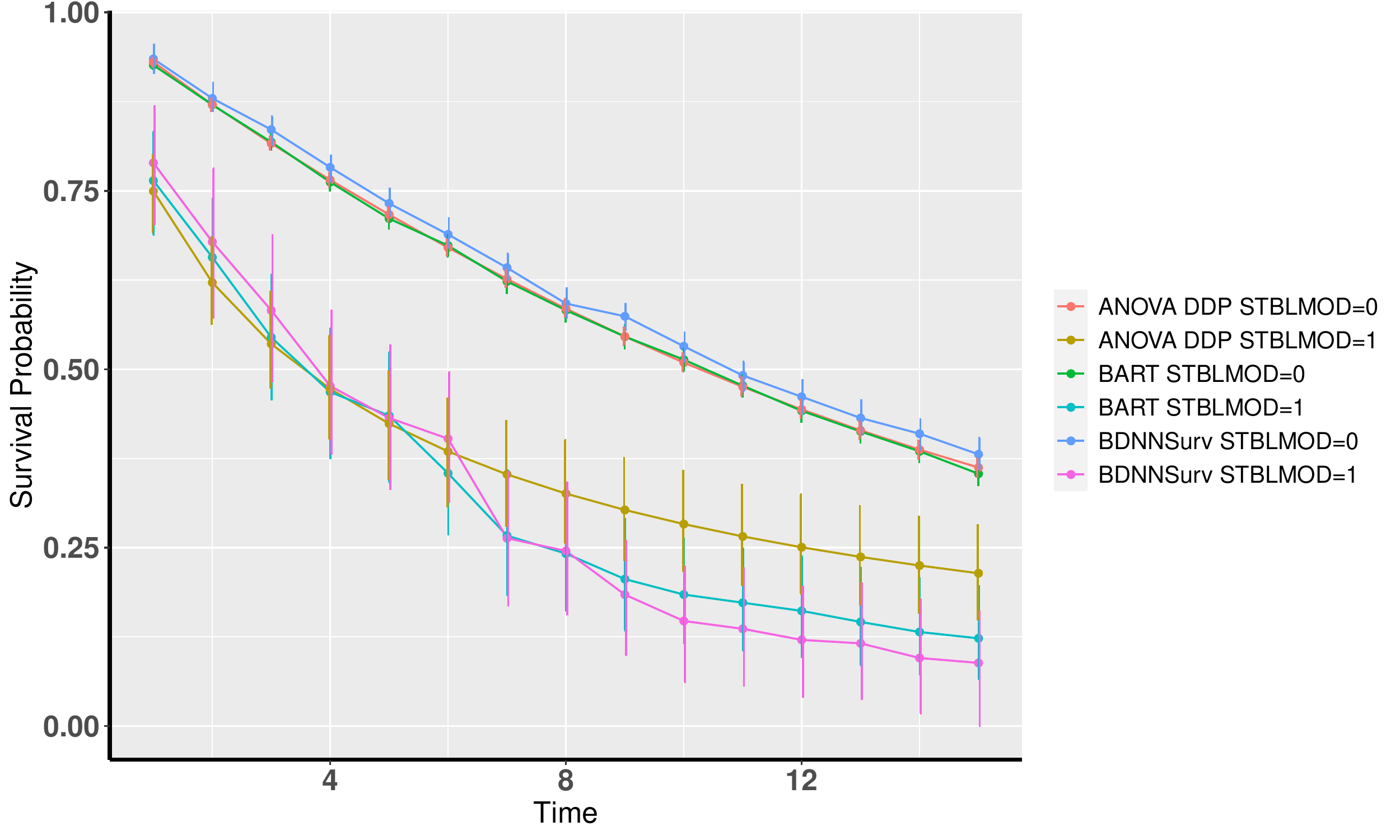}  
  \label{fig:sub-first}
\end{subfigure}
\caption{Estimate of survival probabilities and corresponding credible intervals at different times obtained from using the BDNNSurv, BART, and ANOVA DDP for the CHS data. The three plots present the survival functions stratified by the MI status (MIBLMOD), the CLD status (CHBLMOD), or the stroke status (STBLMOD), respectively.}
\label{fig:chs}
\end{figure}

After building the BDNNSurv model using training data, we predicted the survival probability of subjects in the test data in every year up to 15 years. Figure \ref{fig:chs} shows the survival functions for subjects with certain predictor variables, including 
myocardial infarction, chronic liver diseases, and stroke status, which are known to be important risk factors for the CVD event. The apparent separation of 95\% confidence intervals of survival curves across time points demonstrated that subjects without risk factor (level 0) had significantly larger (at the significance level of 0.05) survival rates as compared to that with risk factor (level 1). 

In the CHS data analysis, to obtain the estimate of survival probabilities at 15 different time points (year 1 to year 15 with an increment of 1 year), our BDNNSurv method is faster compared to the BART and ANOVA DDP. It took about 7, 20, and 185 minutes for BDNNSurv, ANOVA DDP, and BART, respectively, to run 10,000 iterations on an Intel(R) Core(TM) i7-4790 CPU 3.6 GHz, 32 GB RAM computer. 

\section{Concluding Remarks}
\label{sec:final}
Modeling and prediction of TTE data with high-dimensional covariates has recently drawn a lot of attention. Different DNN-based approaches have been studied. The previously proposed methods, however, can only provide point estimate without quantification of the corresponding uncertainty, which can be of crucial importance in predictive  modeling and subsequent decision making.
In this study,  we  proposed the BDNNSurv, a Bayesian  hierarchical  deep  neural  networks  model,  for modeling and prediction of TTE data.  Our proposed model accurately predicted the survival probability along with the corresponding uncertainty,  as demonstrated by simulation studies and real data analysis. 


We specified the prior distributions based on the online documentation of  PyMC3 (\cite{salvatier2016}). For the prior distributions of $W^c_l$ and $b^c_l$, we assumed $W^c_l \sim N(\mu_{W_l}, \sigma_{W_l}^2)$
and $b_l^c \sim N(\mu_{b_l}, \sigma_{b_l}^2)$. We started with  $\mu_{W_l}=\mu_{b_l}=0$ and $\sigma_{W_l}^2=\sigma_{b_l}^2=1$ and also tried $\sigma_{W_l}^2=\sigma_{b_l}^2=0.5$ or 2.
For the prior distribution of $\sigma$, we tried $U(0, 0.2)$, $U(0, 0.1)$, and $U(0, 0.5)$. We decided to use $\sigma_{W_l}^2=\sigma_{b_l}^2=1$ and $U(0, 0.2)$, which produced estimates with good performance, including smaller bias, MSE, and coverage probabilities closer to the nominal value in case 1 of the simulation study. The choice of prior distributions is important in Bayesian neural networks and can have a large influence on the final results \citep{silvestro2020}. We suggest treating these hyper-parameters in the prior distributions as tuning parameters in the data analysis.

In future work, we will extend the proposed framework to accommodate complex biological information, such as gene-expression data, in survival prediction using different architectures e.g. convolutional neural networks \citep{lopez2020}. Moreover, we adopted variational inference to estimate posterior survival probabilities. Instead, in principle, one can compute an approximation to the posterior distribution by sampling multiple predictions with dropout. The dropout approach can also quantify uncertainty in deep learning \citep{gal2016}. In the future, it would be interesting to compare the results from two different methods. 

\beginsupplement
\section*{Supplementary Material}
\label{supp:A}
The code used for simulation, including R code to simulate data and summarize results,
python code to generate initial values for BDNNSurv, python code to run BDNNSurv, and R code to run BART and ANOVA DDP.

\section*{Acknowledgments}
This Manuscript was prepared using the CHS data obtained from the National Heart, Lung,
and Blood Institute (NHLBI) Biologic Specimen and Data Repository Information Coordinating Center (\url{https://biolincc.nhlbi.nih.gov/}), and does not necessarily reflect the opinions or views of the NHLBI. 



\bibliographystyle{jds}
\bibliography{bdnnsurv}

\end{document}